\documentclass[twocolumn,english,pra]{revtex4}
\usepackage[T1]{fontenc}
\usepackage[latin9]{inputenc}
\usepackage{amsmath}
\usepackage{amssymb}
\usepackage{graphicx}
\usepackage{esint}

\makeatletter
\@ifundefined{textcolor}{}
{%
 \definecolor{BLACK}{gray}{0}
 \definecolor{WHITE}{gray}{1}
 \definecolor{RED}{rgb}{1,0,0}
 \definecolor{GREEN}{rgb}{0,1,0}
 \definecolor{BLUE}{rgb}{0,0,1}
 \definecolor{CYAN}{cmyk}{1,0,0,0}
 \definecolor{MAGENTA}{cmyk}{0,1,0,0}
 \definecolor{YELLOW}{cmyk}{0,0,1,0}
}

\makeatother

\usepackage{babel}
\begin{document}

\title{Momentum-resolved radio-frequency spectroscopy of a spin-orbit coupled
atomic Fermi gas near a Feshbach resonance in harmonic traps}

\author{Shi-Guo Peng$^{1}$, Xia-Ji Liu$^{2}$, Hui Hu$^{2}$, and Kaijun
Jiang$^{1,3}$}

\email{kjjiang@wipm.ac.cn}

\affiliation{$^{1}$State Key Laboratory of Magnetic Resonance and Atomic and
Molecular Physics, Wuhan Institute of Physics and Mathematics, Chinese
Academy of Sciences, Wuhan 430071, China }

\affiliation{$^{2}$ARC Centre of Excellence for Quantum-Atom Optics, Centre for
Atom Optics and Ultrafast Spectroscopy, Swinburne University of Technology,
Melbourne 3122, Australia}

\affiliation{$^{3}$Center for Cold Atom Physics, Chinese Academy of Sciences,
Wuhan 430071, China}

\date{\today}
\begin{abstract}
We theoretically investigate the momentum-resolved radio-frequency
spectroscopy of a harmonically trapped atomic Fermi gas near a Feshbach
resonance in the presence of equal Rashba and Dresselhaus spin-orbit
coupling. The system is qualitatively modeled as an ideal gas mixture
of atoms and molecules, in which the properties of molecules, such
as the wavefunction, binding energy and effective mass, are determined
from the two-particle solution of two-interacting atoms. We calculate
separately the radio-frequency response from atoms and molecules at
finite temperatures by using the standard Fermi golden rule, and take
into account the effect of harmonic traps within local density approximation.
The total radio-frequency spectroscopy is discussed, as functions
of temperature and spin-orbit coupling strength. Our results give
a qualitative picture of radio-frequency spectroscopy of a resonantly
interacting spin-orbit coupled Fermi gas and can be directly tested
in atomic Fermi gases of $^{40}$K atoms at Shanxi University and
of $^{6}$Li atoms at MIT.
\end{abstract}


\maketitle

\section{introduction}

Thanks to the high-controllability of ultracold atoms in interatomic
interaction, geometry, purity, atomic species and lattice constant
(of optical lattices), ultracold atomic gases have already become
one of the most important footstones in modern physics, and can be
used in simulating various strongly correlated many-body models in
solid state physics \cite{Bloch2008M,Giorgini2008T,Chin2010F}. Very
recently, another controllability of ultracold atoms is realized experimentally,
in which the spin degree of freedom of atoms (i.e., the atomic internal
hyperfine states) is coupled to the orbital degree of freedom (i.e.,
the momentum) by using a two-photon Raman process. This so-called
spin-orbit (SO) coupling was first created and detected in atomic
$^{87}$Rb Bose-Einstein condensates (BEC) \cite{Lin2011A,Lin2011S},
and then produced in atomic Fermi gases of $^{40}$K atoms \cite{Wang2012S}
and $^{6}$Li atoms \cite{Cheuk2012S}. The realization of SO coupled
atomic gases enables the simulation of charged particles by using
neutral atoms, which are cleaner and more controllable \cite{Dalibard2011A}.
These experiments open an entirely new way to study the celebrated
effects of SO coupling, such as topological insulators, topological
superconductors and new exotic superfluid phases \cite{Qi2010T,Hasan2010T,Zhai2012S}.

Radio-frequency (rf) spectroscopy has been widely applied in many
experiments to study fermionic pairing in a two-component Fermi gas
near Feshbach resonances when it crosses from a Bardeen-Cooper-Schrieffer
(BCS) superfluid of weakly interacting Cooper pairs over to a BEC
of tightly bound molecules \cite{Chin2004O,Schunck2008D}, and also
to investigate the properties of polarons in spin-imbalanced Fermi
gases \cite{Schirotzek2009O,Zhang2012P,Kohstall2012M,Koschorreck2012A}.
This technology allows a precise determination of the molecular binding
energy \cite{Chin2004O,Regal2003C} and pairing gap in a degenerate
Fermi gas \cite{Chin2004O}. In addition, the momentum-resolved rf
spectroscopy, i.e., the spectroscopy at a specific momentum, is also
available \cite{Stewart2008U} and provides important information
on low-energy excitations of novel exotic state of matter in ultracold
atoms. In this respect, it would be a powerful tool for characterizing
the recently realized spin-orbit coupled atomic Fermi gases. Indeed,
momentum-resolved rf spectroscopy of \emph{non-interacting} spin-orbit
coupled $^{40}$K and $^{6}$Li atoms has already been reported \cite{Wang2012S,Cheuk2012S}.

In this work, we aim to qualitatively predict the momentum-resolved
rf spectroscopy of a \emph{resonantly interacting} atomic Fermi gas
with equal Rashba and Dresselhaus spin-orbit coupling, a system already
realized experimentally at Shanxi University and also at MIT by using
broad Feshbach resonances \cite{ZhangPrivateComm,ZwierleinPrivateComm}.
As is well-known, a strongly interacting Fermi gas near Feshbach resonances
is notoriously difficult to handle theoretically \cite{Hu2008,Hu2010}.
Therefore, at the \emph{qualitative} level, we approximate the strongly
interacting Fermi gas as a mixture of non-interacting atoms and molecules.
All the properties of individual molecules are determined from the
two-particle solution of two-interacting atoms. Our approximation
may be justified at large temperatures well above the superfluid transition
temperature $T_{c}$, where molecules are formed below the Feshbach
resonance and have little correlations among themselves or with atoms.
This is exactly the situation in current experiments, for instance,
at Shanxi University \cite{Wang2012S}, the typical temperature of
spin-orbit coupled $^{40}$K atoms is now at about $0.6T_{F}$, where
$T_{F}$ is the Fermi degenerate temperature.

We consider separately the momentum-resolved rf-spectroscopy of atoms
and of molecules. Moreover, according to real experiments we take
into account the crucial trapping potential that prevents the atoms
and molecules from escaping. Within local density approximation (LDA),
the trapped Fermi cloud may be divided into many locally uniform subsystems.
For each subsystem, we calculate the rf responses of the atomic and
molecular components based on the one-particle and two-particle solutions
of uniform systems and Fermi's golden rule. We finally add up all
the local contributions to determine the total momentum-resolved rf
spectroscopy. We note that due to SO coupling, weakly bound molecules
with anisotropic mass and wavefunction may be formed \cite{Vyasanakere2011B,Hu2011P,Yu2011S}.
The bound to free rf transition of weakly bound molecules at rest
has been predicted \cite{Hu2012R}. In this work, we will consider
the rf transition of weakly bound molecules at motion with arbitrary
center-of-mass (COM) momentum.

The paper is arranged as follows. In the next section, we consider
the situation of current experiments, and introduce the model Hamiltonian
responsible for equal Rashba and Dresselhaus SO coupling. In addition,
we introduce the LDA formulism and present the calculation for the
chemical potential. Then, the single-particle and two-particle problems
of a local uniform subsystem are solved in Sec. \ref{sec:single--and-two-body}.
The general formulas for the binding energy and wavefunction of two-particle
bound state with non-zero COM momentum are also provided. In Sec.
\ref{sec:radio-frequency-spectroscopy}, we derive the momentum-resolved
rf transition signals for\emph{ }non-interacting\emph{ }atoms and
molecules, respectively, and then obtain the total rf spectroscopy
for the harmonically trapped\emph{ }ideal\emph{ }gas mixture of fermionic
atoms and bosonic molecules. Finally, our main results are concluded
in Sec.\ref{sec:conclusions}.

\section{models\label{sec:models}}

\subsection{Hamiltonian}

We consider a SO coupled Fermi gas with atomic mass $m$ in a harmonic
trap $V\left(\mathbf{r}\right)=m(\omega_{x}^{2}x^{2}+\omega_{y}^{2}y^{2}+\omega_{z}^{2}z^{2})/2$.
The SO coupling is created by a two-photon Raman process \cite{Wang2012S,Cheuk2012S}.
The many-body Hamiltonian responsible for this process may be modeled
as $\mathcal{H}=\mathcal{H}_{0}+\mathcal{H}_{int}$, where
\begin{eqnarray}
\mathcal{H}_{0} & = & \sum_{\sigma}\int d\mathbf{r}\Psi_{\sigma}^{\dagger}\left(\mathbf{r}\right)\left[-\frac{\hbar^{2}\mathbf{\nabla}^{2}}{2m}+V\left(\mathbf{r}\right)-\mu\right]\Psi_{\sigma}\left(\mathbf{r}\right)+\nonumber \\
 &  & \int d\mathbf{r}\left[\Psi_{\uparrow}^{\dagger}\left(\mathbf{r}\right)\left(\frac{\Omega_{R}}{2}e^{+i2k_{R}x}\right)\Psi_{\downarrow}\left(\mathbf{r}\right)+H.c.\right]\label{eq:2.1.1}
\end{eqnarray}
 is the single-particle Hamiltonian and
\begin{equation}
\mathcal{H}_{int}=U_{0}\int d\mathbf{r}\Psi_{\uparrow}^{\dagger}\left(\mathbf{r}\right)\Psi_{\downarrow}^{\dagger}\left(\mathbf{r}\right)\Psi_{\downarrow}\left(\mathbf{r}\right)\Psi_{\uparrow}\left(\mathbf{r}\right)\label{eq:2.1.2}
\end{equation}
is the interaction Hamiltonian describing the contact interaction
between two spin states. Here, $\Psi_{\sigma}^{\dagger}\left(\mathbf{r}\right)$
is the creation field operator for atoms in the spin-state $\sigma$.
The second term in $\mathcal{H}_{0}$ describes the Raman coupling
between two spin states with strength $\Omega_{R}$ , $k_{R}$ is
the wave-number of two Raman laser beams and therefore $2\hbar k_{R}$
is the momentum transfer during the two-photon Raman process \cite{Wang2012S,Cheuk2012S}.
The interaction strength is denoted by the bare interaction parameter
$U_{0}$, which should be regularized in terms of the \textit{s}-wave
scattering length $a_{s}$, i.e., $1/U_{0}=m/\left(4\pi\hbar^{2}a_{s}\right)-(1/V)\sum_{\mathbf{k}}m/\left(\hbar^{2}k^{2}\right)$.
By using Feshbach resonances, the \textit{s}-wave scattering length
$a_{s}$ could be arbitrarily tuned, and the system may undergo a
crossover from a BCS superfluid of weakly interacting Cooper pairs
to a BEC of tightly bound molecules. Therefore, the atomic chemical
potential $\mu$ may decrease as a result of decreasing atomic population,
when weakly bound molecules are formed.

As argued in the Introduction section, we will treat the Fermi cloud
as a mixture of non-interacting atoms and molecules, although the
degree of freedom of molecules is not made explicitly in our model
Hamiltonian from the outset. We will take the molecule as a two-particle
bound state and determine all its properties by using two-particle
solution of two interacting atoms. In the harmonic trap, we assume
that these molecules experience the same harmonic trap as atoms, but
with double mass, i.e., $V_{M}\left(\mathbf{r}\right)=m(\omega_{x}^{2}x^{2}+\omega_{y}^{2}y^{2}+\omega_{z}^{2}z^{2})$.
Note that, our treatment of ideal mixture of atoms and molecules becomes
exact for a Feshbach resonance with \emph{zero} resonance width. However,
for the broad Feshbach resonance used experimentally, this treatment
is valid at the qualitative level only.

\subsection{Local density approximation}

If the number of atoms is sufficiently large, it is reasonable to
assume that the trapped cloud may be divided into many locally uniform
subsystems with a local chemical potential \cite{Yi2006T,Liu2007M}.
Then, within LDA, the external trap $V\left(\mathbf{r}\right)$ in
the Hamiltonian (\ref{eq:2.1.1}) is absorbed into the chemical potential,
and we can define \emph{effective} local chemical potential
\begin{equation}
\mu\left(\mathbf{r}\right)=\mu-\frac{1}{2}m(\omega_{x}^{2}x^{2}+\omega_{y}^{2}y^{2}+\omega_{z}^{2}z^{2}).\label{eq:2.2.1}
\end{equation}
Note that the global chemical potential $\mu$, which is determined
by the total number of atoms, can be regarded as the local chemical
potential in the trap center ($\mathbf{r}=0$).

In order to evaluate the chemical potential $\mu$, let us consider
the ideal gas mixture of fermionic atoms and bosonic molecules at
finite temperatures. The distributions of the atoms and molecules
are respectively determined by the Fermi-Dirac and Bose-Einstein distributions,
\begin{equation}
f_{A}\left(\epsilon\right)=\frac{1}{e^{\left(\epsilon-\mu\right)/\left(k_{B}T_{A}\right)}+1},\label{eq:2.2.2}
\end{equation}
\begin{equation}
f_{M}\left(\epsilon\right)=\frac{1}{e^{\left(\epsilon-\mu_{M}\right)/\left(k_{B}T_{M}\right)}-1},\label{eq:2.2.3}
\end{equation}
where $k_{B}$ is Boltzmann's constant, and $T_{A,M}$, $\mu$ and
$\mu_{M}$ are the temperatures, chemical potentials of atoms and
molecules, respectively. From the thermal and chemical equilibrium
conditions, we have the following relations, $T_{A}=T_{M}\equiv T$
and $\mu_{M}=2\mu+\varepsilon_{B}$, where $\varepsilon_{B}$ is the
binding energy of the molecules relative to the threshold of two free
atoms. To ease the numerical workload, we will approximate it as $\varepsilon_{B}\simeq E_{B}\equiv\hbar^{2}/\left(ma_{s}^{2}\right)$.
Here, $E_{B}$ is the binding energy in the absence of SO coupling.

The populations of the atoms $N_{A}$ (in a single spin component)
and \emph{noncondensed} molecules $N_{M}$ are given by
\begin{equation}
N_{A}=\int_{0}^{\infty}d\epsilon\rho_{A}\left(\epsilon\right)f_{A}\left(\epsilon\right),\label{eq:2.2.6}
\end{equation}
\begin{equation}
N_{M}=\int_{0}^{\infty}d\epsilon\rho_{M}\left(\epsilon\right)f_{M}\left(\epsilon\right),\label{eq:2.2.7}
\end{equation}
where $\rho_{A,M}\left(\epsilon\right)$ are the density of states
(DOS) of the atoms and molecules, respectively. Qualitatively, for
atoms we shall use the well-known expression for DOS in harmonically
trapped systems without SO coupling,
\begin{equation}
\rho_{A}\left(\epsilon\right)d\epsilon=\frac{\epsilon^{2}}{2\left(\hbar\omega\right)^{3}}d\epsilon,\label{eq:2.2.8}
\end{equation}
where $\omega\equiv(\omega_{x}\omega_{y}\omega_{z})^{1/3}$. For molecules,
we instead use,
\begin{equation}
\rho_{M}\left(\epsilon\right)d\epsilon=\frac{\sqrt{\gamma}\epsilon^{2}}{2\left(\hbar\omega\right)^{3}}d\epsilon.\label{eq:2.2.9}
\end{equation}
The factor $\sqrt{\gamma}$ appearing in the DOS of molecules is due
to the SO coupling, which induces an effective mass $M_{x}=\gamma\cdot(2m)$
as we will make clear later. Combining Eqs. (\ref{eq:2.2.2}) and
(\ref{eq:2.2.3}) and (\ref{eq:2.2.6})-(\ref{eq:2.2.9}), the populations
of the atoms and \emph{noncondensed} molecules are given by
\begin{equation}
N_{A}=-\left(\frac{k_{B}T}{\hbar\omega}\right)^{3}Li_{3}\left(-z_{A}\right),\label{eq:2.2.10}
\end{equation}
\begin{equation}
N_{M}=\sqrt{\gamma}\left(\frac{k_{B}T}{\hbar\omega}\right)^{3}Li_{3}\left(z_{M}\right),\label{eq:2.2.11}
\end{equation}
where $z_{A}=e^{\mu/\left(k_{B}T\right)}$ and $z_{M}=e^{\left(2\mu+E_{B}\right)/\left(k_{B}T\right)}$
are the fugacities of the atoms and molecules, respectively, and $Li_{n}\left(z\right)$
is the polylogarithm function. Therefore, the chemical potential $\mu$
should satisfy the following equation,
\begin{equation}
2N_{A}+2\left[N_{M}+N_{M}^{(0)}\right]=N,\label{eq:2.2.12}
\end{equation}
where $N_{M}^{(0)}$ is the population of the \emph{condensed} molecules,
and $N$ is the total number of atoms. Note that one molecule is counted
as two atoms. Above the critical temperature, i.e., $T>T_{c}$, there
is no condensed molecules ($N_{M}^{(0)}=0$), therefore Eqs. (\ref{eq:2.2.10})-(\ref{eq:2.2.12})
yield
\begin{equation}
-2Li_{3}\left(-z_{A}\right)+2\sqrt{\gamma}Li_{3}\left(z_{M}\right)=N\left(\frac{\hbar\omega}{k_{B}T}\right)^{3}.\label{eq:2.2.13}
\end{equation}
Then from Eq. (\ref{eq:2.2.13}), the chemical potential $\mu$ can
numerically be evaluated at a specific temperature and interaction
strength $E_{B}$. The molecule BEC occurs below the critical temperature,
i.e., $T<T_{c}$. In this case, the chemical potential of molecules
vanishes, i.e., $\mu_{M}=0$, which in turn gives $\mu=-E_{B}/2$.

With the chemical potential $\mu$ obtained, the properties of the
trapped systems may be calculated by integrating over the whole space,
based on the local solutions of uniform systems. In the following
sections, we will firstly solve single-particle and two-particle problems
in uniform systems, and then investigate the rf spectroscopy of trapped
gases under the LDA approximation.

\section{single- and two-particle problems for uniform systems\label{sec:single--and-two-body}}

\subsection{Single-particle solution}

The single-particle problem in a uniform system has been discussed
in detail in our previous work \cite{Hu2012R,Liu2012M}. Here, for
self-containment we summarize briefly the results. We focus on the
following non-interacting Hamiltonian for the subsystem,
\begin{eqnarray}
\mathcal{H}_{0} & = & \sum_{\sigma}\int d\mathbf{r}\Psi_{\sigma}^{\dagger}\left(\mathbf{r}\right)\frac{\hbar^{2}\mathbf{\hat{k}}^{2}}{2m}\Psi_{\sigma}\left(\mathbf{r}\right)+\nonumber \\
 &  & \frac{\Omega_{R}}{2}\int d\mathbf{r}\left[\Psi_{\uparrow}^{\dagger}\left(\mathbf{r}\right)e^{i2k_{R}x}\Psi_{\downarrow}\left(\mathbf{r}\right)+H.c.\right],\label{eq:3.1.1}
\end{eqnarray}
where $\mathbf{\hat{k}}\equiv-i\nabla$. Under the transformation
\cite{Hu2012R,Liu2012M}, $\Psi_{\uparrow}\left(\mathbf{r}\right)=e^{+ik_{R}x}\psi_{\uparrow}\left(\mathbf{r}\right)$
and $\Psi_{\downarrow}\left(\mathbf{r}\right)=e^{-ik_{R}x}\psi_{\downarrow}\left(\mathbf{r}\right)$,
the spatial dependence of the Raman coupling term in Eq.(\ref{eq:3.1.1})
is removed and yields 
\begin{eqnarray}
\mathcal{H}_{0} & = & \int d\mathbf{r}\psi_{\uparrow}^{\dagger}\left(\mathbf{r}\right)\left[\frac{\hbar^{2}\left(k^{2}+k_{R}^{2}\right)}{2m}+\lambda k_{x}\right]\psi_{\uparrow}\left(\mathbf{r}\right)+\nonumber \\
 &  & \int d\mathbf{r}\psi_{\downarrow}^{\dagger}\left(\mathbf{r}\right)\left[\frac{\hbar^{2}\left(k^{2}+k_{R}^{2}\right)}{2m}-\lambda k_{x}\right]\psi_{\downarrow}\left(\mathbf{r}\right)+\nonumber \\
 &  & h\int d\mathbf{r}\left[\psi_{\uparrow}^{\dagger}\left(\mathbf{r}\right)\psi_{\downarrow}\left(\mathbf{r}\right)+H.c.\right],\label{eq:3.1.4}
\end{eqnarray}
 where $\mathbf{e}_{x}$ is the unit vector along $x$ direction.
For convenience, the effective Zeeman field $h\equiv\Omega_{R}/2$
and the SO coupling strength $\lambda\equiv\hbar^{2}k_{R}/m$ are
introduced. This single-particle Hamiltonian (\ref{eq:3.1.4}) can
be easily diagonalized to yield two eigenvalues \cite{Hu2012R,Liu2012M},
\begin{equation}
E_{\mathbf{k}\pm}=\frac{\hbar^{2}k_{R}^{2}}{2m}+\frac{\hbar^{2}k^{2}}{2m}\pm\sqrt{h^{2}+\lambda^{2}k_{x}^{2}}.\label{eq:3.1.5}
\end{equation}
 The symbols ``$\pm$'' stands for the two helicity branches, and
the corresponding single-atom eigenstates in helicity basis take the
form
\begin{equation}
\left[\begin{array}{c}
\left|\mathbf{k}+\right\rangle \\
\left|\mathbf{k}-\right\rangle 
\end{array}\right]=\left[\begin{array}{cc}
\cos\theta_{\mathbf{k}} & \sin\theta_{\mathbf{k}}\\
-\sin\theta_{\mathbf{k}} & \cos\theta_{\mathbf{k}}
\end{array}\right]\left[\begin{array}{c}
\left|\mathbf{k}\uparrow\right\rangle \\
\left|\mathbf{k}\downarrow\right\rangle 
\end{array}\right],\label{eq:3.1.6}
\end{equation}
 where $\theta_{\mathbf{k}}=\arctan\left[\left(\sqrt{h^{2}+\lambda^{2}k_{x}^{2}}-\lambda k_{x}\right)/h\right]$.
It is obvious that the lowest single-particle energy occurs at $k_{\perp}=\sqrt{k_{y}^{2}+k_{z}^{2}}=0$
and $k_{x}=\sqrt{m^{2}\lambda^{2}/\hbar^{4}-h^{2}/\lambda^{2}}$ if
$h<m\lambda^{2}/\hbar^{2}$, and is given by,
\begin{equation}
E_{min}=\frac{\hbar^{2}k_{R}^{2}}{2m}-\frac{m\lambda^{2}}{2\hbar^{2}}-\frac{\hbar^{2}h^{2}}{2m\lambda^{2}}=-\frac{\hbar^{2}h^{2}}{2m\lambda^{2}}.\label{eq:3.1.7}
\end{equation}

\subsection{Two-body bound state}

The two-body problem with \emph{zero} COM momentum in a uniform system
has already been solved in \cite{Hu2012R}. However, real experiments
are carried out well above the critical temperature \cite{Wang2012S,Cheuk2012S},
where two-particle states with a \emph{non-zero} COM momentum may
become important. As we shall see, at the quantitive level, molecules
with a nonzero COM momentum can have small contribution to the (momentum-resolved)
rf spectroscopy. Here, we outline briefly the general solution with
arbitrary COM momenta. 

Under the gauge transformation for the field operators, the form of
the interatomic interaction (\ref{eq:2.1.2}) is invariant, $\mathcal{H}_{int}=U_{0}\int d\mathbf{r}\psi_{\uparrow}^{\dagger}\left(\mathbf{r}\right)\psi_{\downarrow}^{\dagger}\left(\mathbf{r}\right)\psi_{\downarrow}\left(\mathbf{r}\right)\psi_{\uparrow}\left(\mathbf{r}\right)$.
For a two-body problem in the presence of SO coupling, the COM momentum
is a good quantum number, and therefore the two-body wavefunction
at a given COM momentum $\mathbf{q}$ can generally be written as
(we set the volume $V=1$) \cite{Yu2011S} \begin{widetext}
\begin{multline}
\left|\Psi_{2B}\left(\mathbf{q}\right)\right\rangle =\frac{1}{\sqrt{2\mathcal{C}}}\sum_{\mathbf{k}}\left[\psi_{\uparrow\downarrow}\left(\mathbf{k}\right)c_{\mathbf{q}/2+\mathbf{k}\uparrow}^{\dagger}c_{\mathbf{q}/2-\mathbf{k}\downarrow}^{\dagger}+\psi_{\downarrow\uparrow}\left(\mathbf{r}\right)c_{\mathbf{q}/2+\mathbf{k}\downarrow}^{\dagger}c_{\mathbf{q}/2-\mathbf{k}\uparrow}^{\dagger}+\right.\\
\left.\psi_{\uparrow\uparrow}\left(\mathbf{k}\right)c_{\mathbf{q}/2+\mathbf{k}\uparrow}^{\dagger}c_{\mathbf{q}/2-\mathbf{k}\uparrow}^{\dagger}+\psi_{\downarrow\downarrow}\left(\mathbf{k}\right)c_{\mathbf{q}/2+\mathbf{k}\downarrow}^{\dagger}c_{\mathbf{q}/2-\mathbf{k}\downarrow}^{\dagger}\right]\left|vac\right\rangle ,\label{eq:3.2.2}
\end{multline}
where $c_{\mathbf{k}\uparrow}^{\dagger}$ and $c_{\mathbf{k}\downarrow}^{\dagger}$
are creation field operators of spin-up and spin-down atoms with momentum
$\mathbf{k}$, and $\mathcal{C}=\sum_{{\bf k}}[\left|\psi_{\uparrow\downarrow}\left({\bf k}\right)\right|^{2}+\left|\psi_{\downarrow\uparrow}\left({\bf k}\right)\right|^{2}+\left|\psi_{\uparrow\uparrow}\left({\bf k}\right)\right|^{2}+\left|\psi_{\downarrow\downarrow}\left({\bf k}\right)\right|^{2}]$
is the normalization factor of the two-body wavefunction. The Schr\"{o}dinger
equation $\mathcal{H}\left|\Psi_{2B}\left(\mathbf{q}\right)\right\rangle =\varepsilon_{B}\left(\mathbf{q}\right)\left|\Psi_{2B}\left(\mathbf{q}\right)\right\rangle $
is equivalent to the following set of coupled equations,

\begin{eqnarray}
\left[\varepsilon_{B}\left(\mathbf{q}\right)-\left(\frac{\hbar^{2}k_{R}^{2}}{m}+\frac{\hbar^{2}k^{2}}{m}+\frac{\hbar^{2}q^{2}}{4m}+2\lambda k_{x}\right)\right]\psi_{\uparrow\downarrow}\left({\bf k}\right) & = & +(U_{0}/2)\sum\nolimits _{{\bf k}^{\prime}}\left[\psi_{\uparrow\downarrow}\left({\bf k}^{\prime}\right)-\psi_{\downarrow\uparrow}\left({\bf k}^{\prime}\right)\right]+h\psi_{\uparrow\uparrow}\left({\bf k}\right)+h\psi_{\downarrow\downarrow}\left({\bf k}\right),\label{eq:3.2.4}\\
\left[\varepsilon_{B}\left(\mathbf{q}\right)-\left(\frac{\hbar^{2}k_{R}^{2}}{m}+\frac{\hbar^{2}k^{2}}{m}+\frac{\hbar^{2}q^{2}}{4m}-2\lambda k_{x}\right)\right]\psi_{\downarrow\uparrow}\left({\bf k}\right) & = & -(U_{0}/2)\sum\nolimits _{{\bf k}^{\prime}}\left[\psi_{\uparrow\downarrow}\left({\bf k}^{\prime}\right)-\psi_{\downarrow\uparrow}\left({\bf k}^{\prime}\right)\right]+h\psi_{\uparrow\uparrow}\left({\bf k}\right)+h\psi_{\downarrow\downarrow}\left({\bf k}\right),\label{eq:3.2.5}\\
\left[\varepsilon_{B}\left(\mathbf{q}\right)-\left(\frac{\hbar^{2}k_{R}^{2}}{m}+\frac{\hbar^{2}k^{2}}{m}+\frac{\hbar^{2}q^{2}}{4m}-\lambda q_{x}\right)\right]\psi_{\uparrow\uparrow}\left({\bf k}\right) & = & h\psi_{\uparrow\downarrow}\left({\bf k}\right)+h\psi_{\downarrow\uparrow}\left({\bf k}\right),\label{eq:3.2.6}\\
\left[\varepsilon_{B}\left(\mathbf{q}\right)-\left(\frac{\hbar^{2}k_{R}^{2}}{m}+\frac{\hbar^{2}k^{2}}{m}+\frac{\hbar^{2}q^{2}}{4m}+\lambda q_{x}\right)\right]\psi_{\downarrow\downarrow}\left({\bf k}\right) & = & h\psi_{\uparrow\downarrow}\left({\bf k}\right)+h\psi_{\downarrow\uparrow}\left({\bf k}\right).\label{eq:3.2.7}
\end{eqnarray}
\end{widetext}Here, $\varepsilon_{B}\left(\mathbf{q}\right)$ is
the energy of the two-body wavefunction which is dependent on the
COM momentum $\mathbf{q}$. By introducing
\begin{equation}
A_{{\bf kq}}\equiv\varepsilon_{B}\left(\mathbf{q}\right)-\left(\frac{\hbar^{2}k_{R}^{2}}{m}+\frac{\hbar^{2}k^{2}}{m}+\frac{\hbar^{2}q^{2}}{4m}\right),\label{eq:3.2.8}
\end{equation}
$\psi_{s}\left(\mathbf{k}\right)=[\psi_{\uparrow\downarrow}\left({\bf k}\right)-\psi_{\downarrow\uparrow}\left({\bf k}\right)]/\sqrt{2}$
and $\psi_{a}\left({\bf k}\right)=[\psi_{\uparrow\downarrow}\left({\bf k}\right)+\psi_{\downarrow\uparrow}\left({\bf k}\right)]/\sqrt{2}$,
it is easy to show that,

\begin{eqnarray}
\psi_{\uparrow\uparrow}\left({\bf k}\right) & = & \frac{\sqrt{2}h}{A_{{\bf kq}}-\lambda q_{x}}\psi_{a}\left({\bf k}\right),\label{eq:3.2.11}\\
\psi_{\downarrow\downarrow}\left({\bf k}\right) & = & \frac{\sqrt{2}h}{A_{{\bf kq}}+\lambda q_{x}}\psi_{a}\left({\bf k}\right),\label{eq:3.2.12}\\
\psi_{a}\left({\bf k}\right) & = & \frac{2\lambda k_{x}}{4h^{2}+\lambda^{2}q_{x}^{2}}\times\nonumber \\
 &  & \left(\frac{\lambda^{2}q_{x}^{2}}{A_{{\bf kq}}}+\frac{4h^{2}A_{{\bf kq}}}{A_{{\bf kq}}^{2}-4h^{2}-\lambda^{2}q_{x}^{2}}\right)\psi_{s}\left({\bf k}\right),\label{eq:3.2.13}
\end{eqnarray}
and
\begin{equation}
\left[A_{{\bf kq}}-\frac{4\lambda^{2}k_{x}^{2}/A_{{\bf kq}}}{1-4h^{2}/\left(A_{{\bf kq}}^{2}-\lambda^{2}q_{x}^{2}\right)}\right]\psi_{s}\left({\bf k}\right)=U_{0}\sum_{{\bf k}^{\prime}}\psi_{s}\left({\bf k}^{\prime}\right).\label{eq:3.2.14}
\end{equation}
From Eq. (\ref{eq:3.2.14}), the un-normalized spin-singlet wavefunction
$\psi_{s}\left({\bf k}\right)$ is given by,
\begin{equation}
\psi_{s}\left({\bf k}\right)=\left[A_{{\bf kq}}-\frac{4\lambda^{2}k_{x}^{2}/A_{{\bf kq}}}{1-4h^{2}/\left(A_{{\bf kq}}^{2}-\lambda^{2}q_{x}^{2}\right)}\right]^{-1}.\label{eq:3.2.15}
\end{equation}
 Thus, the two-body binding energy is determined by $U_{0}\sum_{\mathbf{k}}\psi_{s}\left(\mathbf{k}\right)=1$
, or 
\begin{equation}
\frac{m}{4\pi\hbar^{2}a_{s}}-\sum_{{\bf k}}\left[\psi_{s}\left({\bf k}\right)+\frac{m}{\hbar^{2}k^{2}}\right]=0.\label{eq:3.2.16}
\end{equation}
Here, the bare interaction $U_{0}$ has been regularized by the \textit{s}-wave
scattering length $a_{s}$ . A bound state exists if its energy satisfies
$\varepsilon_{B}\left(\mathbf{q}\right)<2E_{min}$, where $E_{min}$
is the lowest single-atom energy given by Eq. (\ref{eq:3.1.7}). The
normalization factor for the wavefunction is given by

\begin{eqnarray}
\mathcal{C} & = & \sum_{\mathbf{k}}\left|\psi_{s}\left({\bf k}\right)\right|^{2}\left\{ 1+\left[\frac{4h^{2}\left(A_{\mathbf{kq}}^{2}+\lambda^{2}q_{x}^{2}\right)}{\left(A_{\mathbf{kq}}^{2}-\lambda^{2}q_{x}^{2}\right)^{2}}+1\right]\times\right.\nonumber \\
 &  & \left.\frac{4\lambda^{2}k_{x}^{2}/A_{{\bf kq}}^{2}}{\left[1-4h^{2}/\left(A_{{\bf kq}}^{2}-\lambda^{2}q_{x}^{2}\right)\right]^{2}}\right\} .\label{eq:3.2.17}
\end{eqnarray}
\begin{figure}
\includegraphics[width=1\columnwidth]{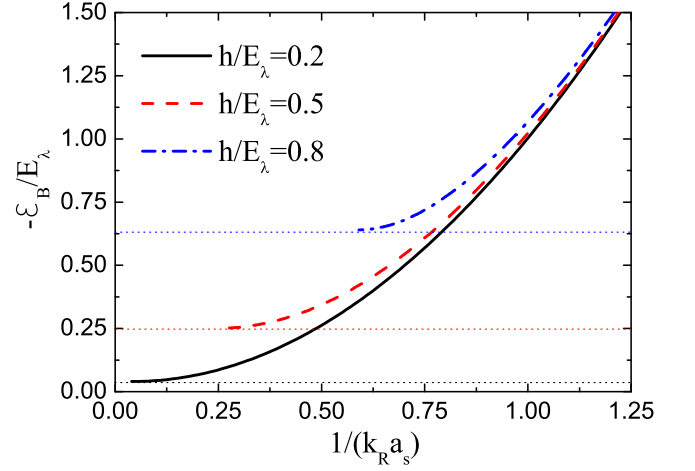}

\caption{(Color online) Binding energies $-\varepsilon_{B}/E_{\lambda}$ of
two atoms with zero COM momentum $\mathbf{q}=0$ as functions of the
\textit{s}-wave scattering length $1/\left(k_{R}a_{s}\right)$ for
three typical values of the Zeeman field. The horizontal dotted lines
are the threshold energies $2E_{min}$ where the bound states disappear.
Here, we have introduced $k_{R}=m\lambda/\hbar^{2}$ and $E_{\lambda}=m\lambda^{2}/\hbar^{2}$
as the units of the wavevector and energy, respectively.}

\label{fig1}
\end{figure}
The bound state energy can be easily solved from Eq. (\ref{eq:3.2.16})
and the results are plotted as a function of the \textit{s}-wave scattering
length in Fig.\ref{fig1} for zero COM momentum at three typical values
of the Zeeman field ($h/E_{\lambda}=0.2,0.5,0.8$). Here, we have
introduced $k_{R}=m\lambda/\hbar^{2}$ and $E_{\lambda}=m\lambda^{2}/\hbar^{2}$
as the units of the wavevector and energy, respectively. We find that
the bound state only exists for $a_{s}>0$.

\begin{figure}
\includegraphics[width=1\columnwidth]{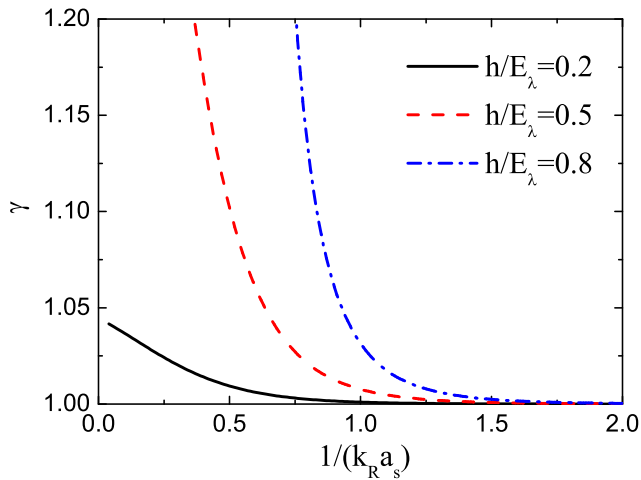}

\caption{(Color online) The effective mass of weakly bound molecules, $\gamma\equiv M_{x}/\left(2m\right)$,
as functions of the interaction strength $1/\left(k_{R}a_{s}\right)$
for three typical values of the Zeeman field. }

\label{fig2}
\end{figure}

Due to the SO coupling, the effective mass of the two-body bound state
is affected. In the limit of small COM momentum, i.e., $\mathbf{q}\approx0$,
the energy of two-body bound state would have a well-defined dispersion,
\begin{equation}
\varepsilon_{B}\left(\mathbf{q}\right)=\varepsilon_{B}\left(0\right)+\frac{\hbar^{2}q_{x}^{2}}{2M_{x}}+\frac{\hbar^{2}q_{\perp}^{2}}{2M_{\perp}},\label{eq:3.2.18}
\end{equation}
where $\varepsilon_{B}\left(0\right)$ is the two-body binding energy
with zero COM momentum, and $M_{x,\perp}$ are the effective masses
in $x$ and transverse ($y$ and $z$) directions, respectively. The
transverse effective mass $M_{\perp}$ is not affected by the SO coupling,
i.e., $M_{\perp}=2m$, since the SO coupling is only applied along
$x$ direction, therefore
\begin{equation}
A_{{\bf kq}}=A_{\mathbf{k}}+\frac{\hbar^{2}q_{x}^{2}}{4m}\left(\frac{2m}{M_{x}}-1\right),\label{eq:3.2.19}
\end{equation}
where $A_{\mathbf{k}}=\varepsilon_{B}\left(0\right)-\hbar^{2}k_{R}^{2}/m-\hbar^{2}k^{2}/m$.
Substituting Eq. (\ref{eq:3.2.19}) into Eq. (\ref{eq:3.2.16}), expanding
up to the second order of $q_{x}$, and comparing the corresponding
coefficients, we can easily obtain
\begin{equation}
\frac{1}{\gamma}\equiv\frac{2m}{M_{x}}=1-\frac{m\lambda^{2}}{\hbar^{2}}\cdot\frac{I_{2}}{I_{1}},\label{eq:3.2.21}
\end{equation}
 where
\begin{eqnarray}
I_{1} & = & -\sum_{\mathbf{k}}\frac{\left(A_{\mathbf{k}}^{2}-4h^{2}\right)^{2}+4\lambda^{2}k_{x}^{2}\left(A_{\mathbf{k}}^{2}+4h^{2}\right)}{4A_{\mathbf{k}}^{2}\left(A_{\mathbf{k}}^{2}-4h^{2}-4\lambda^{2}k_{x}^{2}\right)^{2}},\label{eq:3.2.23}\\
I_{2} & = & \sum_{\mathbf{k}}\frac{16\lambda^{2}h^{2}k_{x}^{2}}{A_{\mathbf{k}}^{3}\left(A_{\mathbf{k}}^{2}-4h^{2}-4\lambda^{2}k_{x}^{2}\right)^{2}}.\label{eq:3.2.24}
\end{eqnarray}
The effective mass ratio $\gamma\equiv M_{x}/\left(2m\right)$ as
a function of the interaction strength is shown in Fig.\ref{fig2}
for three typical values of the Zeeman field ($h/E_{\lambda}=0.2,0.5,0.8$).
We find in the BEC limit, the effective mass is independent of the
values of the Zeeman field, and $\gamma$ approaches unity, since
two atoms are deeply bound and form a tightly bound molecule. When
the interaction is tuned towards the resonance, the two-body bound
state becomes looser, and the SO coupling will induce a relatively
large effective mass. However, the large SO coupling could also destroy
the bound state at a critical \textit{s}-wave scattering length as
we have seen in Fig.\ref{fig1}.

\section{radio-frequency spectroscopy\label{sec:radio-frequency-spectroscopy}}

In this section, we will firstly present the basic idea of the rf
transition based on the Fermi's golden rule, and then derive the Frank-Condon
factor for the rf transition of atoms as well as weakly bound molecules.
Finally, under LDA, we theoretically investigate the rf-spectroscopy
of a harmonically trapped ideal gas mixture of fermionic atoms and
bosonic molecules, which is observable in current experiments.

\subsection{Radio-frequency transition and the Fermi's golden rule}

The basic idear of the rf transition is simple. An rf field is applied
to an atomic Fermi gas with two hyperfine states (denoted as $\left|1\right\rangle =\left|\uparrow\right\rangle $
and $\left|2\right\rangle =\left|\downarrow\right\rangle $), and
drives one of the hyperfine states (i.e., $\left|\downarrow\right\rangle $)
to an upper state $\left|3\right\rangle $ with a bare atomic hyperfine
energy difference $\hbar\omega_{3\downarrow}$ due to the magnetic
field splitting. The Hamiltonian of this rf coupling may be written
as,
\begin{equation}
\mathcal{V}_{rf}=V_{0}\int d\mathbf{r}\left[\psi_{3}^{\dagger}\left(\mathbf{r}\right)\Psi_{\downarrow}\left(\mathbf{r}\right)+H.c.\right],\label{eq:4.1.1}
\end{equation}
 where $\psi_{3}^{\dagger}\left(\mathbf{r}\right)$ creates an atom
in the third state at position $\mathbf{r}$, and $V_{0}$ is the
strength of the rf drive and is related to the Rabi frequency $\omega_{R}$
by $V_{0}=\hbar\omega_{R}/2$. Using the field operator after the
gauge transformation, the rf coupling is 
\begin{equation}
\mathcal{V}_{rf}=V_{0}\int d\mathbf{r}\left[e^{-ik_{R}x}\psi_{3}^{\dagger}\left(\mathbf{r}\right)\psi_{\downarrow}\left(\mathbf{r}\right)+H.c.\right],\label{eq:4.1.2}
\end{equation}
 or in the momentum space,
\begin{equation}
\mathcal{V}_{rf}=V_{0}\sum_{\mathbf{k}}\left(c_{\mathbf{k}-k_{R}\mathbf{e}_{x},3}^{\dagger}c_{\mathbf{k}\downarrow}+H.c.\right)\label{eq:4.1.3}
\end{equation}
 for later convenience. Note that, because of the transformation,
effectively there is a momentum loss of $k_{R}\mathbf{e}_{x}$ for
the transferred atoms.

According to the Fermi's golden rule, the general strength of the
rf transition is proportional to the Franck-Condon factor \cite{Chin2005R},
\begin{equation}
\Gamma\left(\Omega\right)=\left|\left\langle \psi_{f}\right|\mathcal{V}_{rf}\left|\psi_{i}\right\rangle \right|^{2}\delta\left(\Omega-\frac{E_{f}-E_{i}}{\hbar}\right),\label{eq:4.1.4}
\end{equation}
where $\left|\psi_{i,f}\right\rangle $ denote the initial and final
states with corresponding energy $E_{i,f}$ , respectively, and $\Omega$
is the frequency of the rf field. The delta function keeps the energy
conserved during the transition.

\subsection{The single-particle rf transition}

In the single-particle picture, the eigenstates of a SO coupled atom
are the helicity states $\left|\mathbf{k}+\right\rangle $ ($\left|\mathbf{k}-\right\rangle $),
other than the original spin states $\left|\mathbf{k}\uparrow\right\rangle $
($\left|\mathbf{k}\downarrow\right\rangle $). If an atom is initially
prepared in the state $\left|\mathbf{k}-\right\rangle $ with energy
$E_{\mathbf{k}-}$, the rf photon will transfer this atom to a third
empty state $\left|3\right\rangle $. In order to obtain the final
state of the rf transition, it's useful to calculate $\mathcal{V}_{rf}\left|\mathbf{k}-\right\rangle $.
Using Eqs. (\ref{eq:3.1.6}) and (\ref{eq:4.1.3}), we can easily
obtain
\begin{equation}
\mathcal{V}_{rf}\left|\mathbf{k}-\right\rangle =V_{0}\cdot\cos\theta_{\mathbf{k}}c_{\mathbf{k}-k_{R}\mathbf{e}_{x},3}^{\dagger}\left|vac\right\rangle .\label{eq:4.2.1}
\end{equation}
This means after the rf transition, there is a probability of $\left|\cos\theta_{\mathbf{k}}\right|^{2}$
that transfers the atom to the final state $\left|3\right\rangle $
with a momentum $\mathbf{k}-k_{R}\mathbf{e}_{x}$. Similarly, the
final state should be
\begin{equation}
\mathcal{V}_{rf}\left|\mathbf{k}+\right\rangle =V_{0}\cdot\sin\theta_{\mathbf{k}}c_{\mathbf{k}-k_{R}\mathbf{e}_{x},3}^{\dagger}\left|vac\right\rangle ,\label{eq:4.2.2}
\end{equation}
if the atom occupies the state $\left|\mathbf{k}+\right\rangle $
before the rf transition. It also means the probability that transfers
an atom to the final state $\left|3\right\rangle $ with a momentum
$\mathbf{k}-k_{R}\mathbf{e}_{x}$ is $\left|\sin\theta_{\mathbf{k}}\right|^{2}$.
Generally, the atoms are initially prepared in the mixture states
of $\left|\mathbf{k}-\right\rangle $ and $\left|\mathbf{k}+\right\rangle $
with the probability $f\left(E_{\mathbf{k}\pm}-\mu\right)$, respectively,
where $f\left(E_{\mathbf{k}\pm}-\mu\right)$ is the Fermi-Dirac distribution,
and $\mu$ is the chemical potential. According to the Fermi's golden
rule, the total rf transition strength therefore is given by ($V_{0}=1$)
\begin{eqnarray}
\Gamma_{A}\left(\Omega\right) & = & \sum_{\mathbf{k}}\sin^{2}\theta_{\mathbf{k}}f\left(E_{\mathbf{k}+}-\mu\right)\delta\left(\Omega-\frac{\Delta\epsilon_{+}}{\hbar}\right)+\nonumber \\
 &  & \sum_{\mathbf{k}}\cos^{2}\theta_{\mathbf{k}}f\left(E_{\mathbf{k}-}-\mu\right)\delta\left(\Omega-\frac{\Delta\epsilon_{-}}{\hbar}\right),\label{eq:4.2.3}
\end{eqnarray}
 where 
\begin{eqnarray}
\Delta\epsilon_{\pm} & = & \hbar\omega_{3\downarrow}+\frac{\hbar^{2}\left(\mathbf{k}-k_{R}\mathbf{e}_{x}\right)^{2}}{2m}-E_{\mathbf{k}\pm}\nonumber \\
 & = & -\lambda k_{x}\mp\sqrt{h^{2}+\lambda^{2}k_{x}^{2}}\label{eq:4.2.4}
\end{eqnarray}
 is the probable energy difference between the final and initial states.
Without any confusion, we have ignored the bare hyperfine splitting,
i.e., $\omega_{3\downarrow}=0$.

\subsection{Two-body bound to free radio-frequency transition}

The picture of the two-body bound to free rf transition is that the
rf photon breaks a weakly bound molecule and transfers one of the
two atoms to the third state $\left|3\right\rangle $. The final state
is determined by \cite{Hu2012R}
\begin{equation}
\mathcal{V}_{rf}\left|\Psi_{2B}\left(\mathbf{q}\right)\right\rangle =V_{0}\sum_{\mathbf{k}}c_{\mathbf{k}-k_{R}\mathbf{e}_{x},3}^{\dagger}c_{\mathbf{k}^{\prime}\downarrow}\left|\Psi_{2B}\left(\mathbf{q}\right)\right\rangle .\label{eq:4.3.1}
\end{equation}
 Substituting the wavefunction of the two-body bound state (\ref{eq:3.2.2})
into Eq. (\ref{eq:4.3.1}), and after some straightforward algebra,
we shall easily obtain
\begin{multline}
\mathcal{V}_{rf}\left|\Psi_{2B}\left(\mathbf{q}\right)\right\rangle =-\sqrt{\frac{1}{\mathcal{C}}}V_{0}\sum_{\mathbf{k}}c_{\mathbf{q}/2-\mathbf{k}-k_{R}\mathbf{e}_{x},3}^{\dagger}\times\\
\left\{ \left[\psi_{s}\left(\mathbf{k}\right)+\psi_{a}\left(\mathbf{k}\right)\right]c_{\mathbf{q}/2+\mathbf{k}\uparrow}^{\dagger}+\sqrt{2}\psi_{\downarrow\downarrow}\left(\mathbf{k}\right)c_{\mathbf{q}/2+\mathbf{k}\downarrow}^{\dagger}\right\} \left|vac\right\rangle .\label{eq:4.3.2}
\end{multline}
 Using Eq. (\ref{eq:3.1.6}), the final state can be written in terms
of the helicity basis as
\begin{multline}
\mathcal{V}_{rf}\left|\Psi_{2B}\left(\mathbf{q}\right)\right\rangle =-\sqrt{\frac{1}{\mathcal{C}}}V_{0}\sum_{\mathbf{k}}c_{\mathbf{q}/2-\mathbf{k}-k_{R}\mathbf{e}_{x},3}^{\dagger}\times\\
\left(s_{\mathbf{q}/2+\mathbf{k}+}c_{\mathbf{q}/2+\mathbf{k}+}^{\dagger}-s_{\mathbf{q}/2+\mathbf{k}-}c_{\mathbf{q}/2+\mathbf{k}-}^{\dagger}\right)\left|vac\right\rangle ,\label{eq:4.3.3}
\end{multline}
where
\begin{eqnarray}
s_{\mathbf{q}/2+\mathbf{k}+} & = & \left[\psi_{s}\left(\mathbf{k}\right)+\psi_{a}\left(\mathbf{k}\right)\right]\cos\theta_{\mathbf{q}/2+\mathbf{k}}+\nonumber \\
 &  & \sqrt{2}\psi_{\downarrow\downarrow}\left(\mathbf{k}\right)\sin\theta_{\mathbf{q}/2+\mathbf{k}},\label{eq:4.3.4}
\end{eqnarray}
\begin{eqnarray}
s_{\mathbf{q}/2+\mathbf{k}-} & = & \left[\psi_{s}\left(\mathbf{k}\right)+\psi_{a}\left(\mathbf{k}\right)\right]\sin\theta_{\mathbf{q}/2+\mathbf{k}}-\nonumber \\
 &  & \sqrt{2}\psi_{\downarrow\downarrow}\left(\mathbf{k}\right)\cos\theta_{\mathbf{q}/2+\mathbf{k}}.\label{eq:4.3.5}
\end{eqnarray}
Eq. (\ref{eq:4.3.3}) could be understood as follows. During the rf
transition, a molecule with COM momentum $\mathbf{q}$ is broken by
absorbing the rf photon, and then the spin-down atom is transferred
to the third state. Finally, there are two possibilities.  We may
have two atoms in the third state and upper helicity state, respectively,
with a probability $\left|s_{\mathbf{q}/2+\mathbf{k}+}\right|^{2}/\mathcal{C}$.
Also, we may have a probability of $\left|s_{\mathbf{q}/2+\mathbf{k}-}\right|^{2}/\mathcal{C}$
for having two atoms in the third state and lower helicity state,
respectively. Taking into account these two final states, we should
have the following transfer strength or the Franck-Condon factor,
according to the Fermi's golden rule ($V_{0}=1$),

\begin{eqnarray}
\Gamma_{M}\left(\Omega\right) & = & \frac{1}{\mathcal{C}}\sum_{\mathbf{k}}\left|s_{\mathbf{q}/2+\mathbf{k}+}\right|^{2}\delta\left(\Omega-\frac{\Delta\varepsilon_{+}}{\hbar}\right)+\nonumber \\
 &  & \frac{1}{\mathcal{C}}\sum_{\mathbf{k}}\left|s_{\mathbf{q}/2+\mathbf{k}-}\right|^{2}\delta\left(\Omega-\frac{\Delta\varepsilon_{-}}{\hbar}\right),\label{eq:4.3.6}
\end{eqnarray}
 where 
\begin{equation}
\Delta\varepsilon_{\pm}=\frac{\hbar^{2}\left(\mathbf{q}/2-\mathbf{k}-k_{R}\mathbf{e}_{x}\right)^{2}}{2m}+E_{\mathbf{q}/2+\mathbf{k},\pm}-\varepsilon_{B}\left(\mathbf{q}\right)\label{eq:4.3.7}
\end{equation}
 is the probable energy difference between the final and initial states.
Here, we have already ignored the bare hyperfine splitting, i.e.,
$\omega_{3\downarrow}=0$ as well. Eqs.(\ref{eq:4.3.6}) and (\ref{eq:4.3.7})
recover the results of \cite{Hu2012R} when $\mathbf{q}=0$ .

\subsection{Rf spectroscopy for a harmonically trapped\emph{ }gas mixture}

In this subsection, we are ready to investigate theoretically the
rf response signals of the \emph{harmonically trapped} \emph{ideal}
gas mixture of the atoms and molecules, as well as the total rf spectroscopy.

\subsubsection{The signals of the atomic component}

The rf transition strength for atoms in a uniform system is determined
by Eq. (\ref{eq:4.2.3}). In the harmonic trap, the rf transition
strength of atoms at the position $\mathbf{r}$ should be $\Gamma_{A}\left(\Omega,\mathbf{r}\right)$
, in which we shall use the local chemical potential,
\begin{equation}
\mu\left(\mathbf{r}\right)=\mu-\frac{m}{2}(\omega_{x}^{2}x^{2}+\omega_{y}^{2}y^{2}+\omega_{z}^{2}z^{2}),\label{eq:4.4.13}
\end{equation}
where $\mu$ is the chemical potential in the trap center determined
by Eq. (\ref{eq:2.2.13}). Thus, using LDA the total rf transition
strength of the atomic component in a harmonic trap is $\Gamma_{A}\left(\Omega\right)=\int d\mathbf{r}\Gamma_{A}\left(\Omega,\mathbf{r}\right)$. 

\begin{figure}
\includegraphics[width=1\columnwidth]{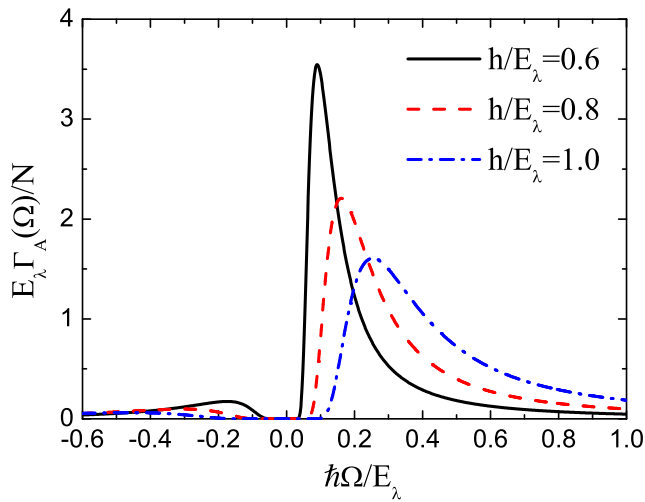}

\caption{(Color online) The rf spectroscopy of the atomic component in a harmonic
trap at different values of the Zeeman field, i.e., $h/E_{\lambda}=0.6,0.8,1.0$.
Here, we have taken $E_{B}/E_{\lambda}=2$, $E_{F}/E_{\lambda}=1$,
and $T/T_{F}=1$. }

\label{fig3}
\end{figure}

In the actual calculation, the summation $\sum_{\mathbf{k}}$ in $\Gamma_{A}\left(\Omega,\mathbf{r}\right)$
may be replaced by the integral $\left(2\pi\right)^{-3}\int dk_{x}\int d^{2}\mathbf{k}_{\perp}$.
After some tedious but straightforward derivation, the total atomic
rf transition strength has the following form,
\begin{eqnarray}
\frac{E_{\lambda}\Gamma_{A}\left(\Omega\right)}{N} & = & -\frac{3}{2\sqrt{2\pi}}\cdot\left(\frac{E_{\lambda}}{E_{F}}\cdot\frac{T}{T_{F}}\right)^{3/2}\frac{h^{2}}{\Omega^{2}}\times\nonumber \\
 &  & \int_{0}^{\infty}dk_{\perp}^{2}Li_{3/2}\left[-z_{A}e^{-E_{k_{\perp}}/\left(k_{B}T\right)}\right],\label{eq:4.4.14}
\end{eqnarray}
 where
\begin{equation}
E_{k_{\perp}}=\frac{\hbar^{2}\left(k_{\perp}^{2}+k_{R}^{2}\right)}{2m}+\frac{\hbar^{2}}{2m\lambda^{2}}\left(\frac{h^{2}-\Omega^{2}}{2\Omega}\right)^{2}-\frac{h^{2}+\Omega^{2}}{2\Omega},\label{eq:4.4.15}
\end{equation}
and $E_{F}=\hbar\omega\left(3N\right)^{1/3}$ and $T_{F}=E_{F}/k_{B}$
are the Fermi energy and Fermi temperature, respectively. At a given
temperature $T$ and interaction strength $E_{B}$, the atomic chemical
potential $\mu$ is obtained from Eq. (\ref{eq:2.2.13}). Then, with
a fixed total atom number $N$ (which in turn gives the Fermi energy
$E_{F}$), the rf transition strength of the atomic component can
numerically be calculated from Eq. (\ref{eq:4.4.14}). The results
are presented in Fig. \ref{fig3} for different values of Zeeman field
at the given parameters, $E_{B}/E_{\lambda}=2$, $E_{F}/E_{\lambda}=1$
and $T/T_{F}=1$. We find that there is a bimodal structure in the
rf spectroscopy of the atomic component. The stronger rf signals at
the positive frequency $\Omega$ correspond to the transition out
of the lower helicity state, while those with weaker strength at the
negative frequency correspond to the transition out of the upper helicity
state. On the other hand, there is a red-shift in the peak positions
as the strength of the Zeeman field decreases. As we may anticipate,
the peak transition frequency $\Omega$ will reach the bare hyperfine
splitting $\omega_{3\downarrow}=0$ in the absence of SO coupling.

In addition to the total strength of the atomic rf transition, we
can also calculate the transition strength of atoms at a specific
momentum $k_{x}$. This momentum-resolved rf spectroscopy has already
been investigated at Shanxi University \cite{Wang2012S} and at MIT
\cite{Cheuk2012S}. According to Eq. (\ref{eq:4.2.3}), we can carry
out the integral over $\mathbf{k}_{\perp}$ and retain that of $k_{x}$.
Therefore, the momentum-resolved rf transition strength is determined
by \begin{widetext}
\begin{eqnarray}
\frac{k_{R}E_{\lambda}\Gamma_{A}\left(\Omega,k_{x}\right)}{N} & = & -\frac{3\sqrt{\pi}}{2\sqrt{2}}\left(\frac{E_{\lambda}}{E_{F}}\cdot\frac{T}{T_{F}}\right)^{3/2}\left(1-\frac{\lambda k_{x}}{\sqrt{h^{2}+\lambda^{2}k_{x}^{2}}}\right)\delta\left(\Omega-\frac{\Delta\epsilon_{+}}{\hbar}\right)\int_{0}^{\infty}dk_{\perp}^{2}Li_{3/2}\left(-z_{A}e^{-E_{\mathbf{k}+}/k_{B}T}\right)+\nonumber \\
 &  & -\frac{3\sqrt{\pi}}{2\sqrt{2}}\left(\frac{E_{\lambda}}{E_{F}}\cdot\frac{T}{T_{F}}\right)^{3/2}\left(1+\frac{\lambda k_{x}}{\sqrt{h^{2}+\lambda^{2}k_{x}^{2}}}\right)\delta\left(\Omega-\frac{\Delta\epsilon_{-}}{\hbar}\right)\int_{0}^{\infty}dk_{\perp}^{2}Li_{3/2}\left(-z_{A}e^{-E_{\mathbf{k}-}/k_{B}T}\right).\label{eq:4.4.16}
\end{eqnarray}
\end{widetext} Corresponding to experiments, the delta function in
Eq. (\ref{eq:4.4.16}) can be replaced by the Lorentz-line shape as,
\begin{equation}
\delta\left(x\right)\rightarrow\frac{1}{\pi}\cdot\frac{\sigma}{x^{2}+\sigma^{2}},\label{eq:4.4.17}
\end{equation}
where $\sigma$ is the line width.

\begin{figure}
\includegraphics[width=1\columnwidth]{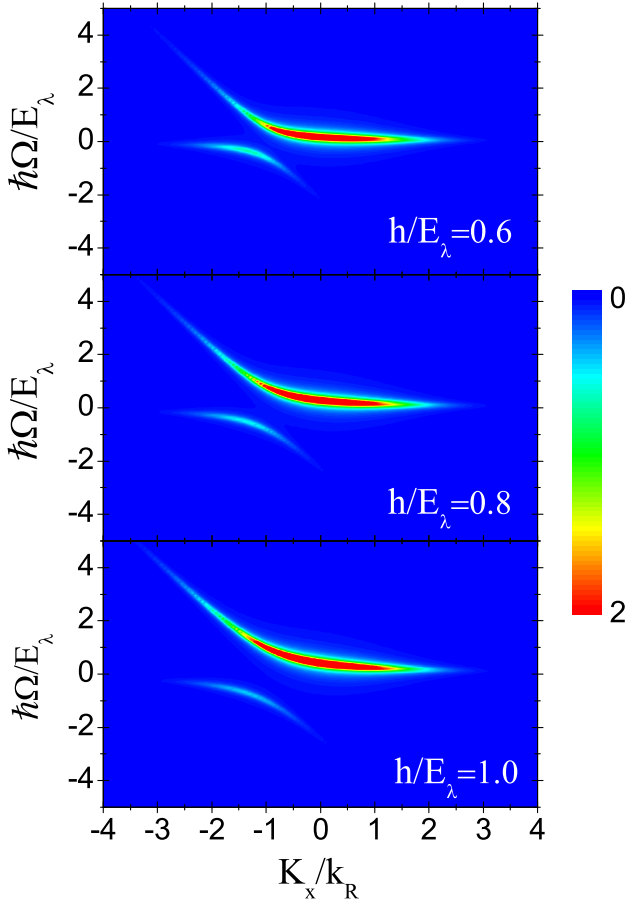}

\caption{(Color online) The momentum-resolved rf spectroscopy of the harmonically
trapped atomic component for different values of the Zeeman field
and interaction strength. Here, we have taken $E_{F}/E_{\lambda}=2$,
$T/T_{F}=1$ and $\sigma=0.1$. In addition, we have also defined
$K_{x}=k_{x}-k_{R}$ (see text), which moves the whole spectra to
the right by an amount $k_{R}$.}

\label{fig4}
\end{figure}

Fig. \ref{fig4} reports the momentum-resolved rf spectroscopy of
the atomic component for different values of the Zeeman field and
interaction strength. Here, we define $K_{x}=k_{x}-k_{R}$ to make
sure the spectra is a symmetric function of $K_{x}$ in the absence
of SO coulping \cite{Liu2012M}. As we anticipate, there are two branches
in which one is relatively weaker than the other one. Since the energy
of the upper helicity state $\left|\mathbf{k}+\right\rangle $ is
larger than that of the lower helicity state $\left|\mathbf{k}-\right\rangle $
, the atoms will occupy the lower helicity state first and then the
upper one before the rf transition. Thus, the initial atomic population
of the lower helicity state should be larger. Therefore, similar to
the integrated rf spectroscopy, the brighter branch of the momentum-resolved
rf spectroscopy corresponds to the rf transition out of the state
$\left|\mathbf{k}-\right\rangle $, while the weaker one corresponds
to the rf transition out of the state $\left|\mathbf{k}+\right\rangle $.
Therefore, the contribution from two initial states could be well
identified experimentally. We can also see that the gap between two
branches becomes larger when the strength of the Zeeman field increases.
This is because the transition frequency deviates from the bare hyperfine
splitting $\omega_{3\downarrow}$ more obviously with a stronger SO
coupling. 

All these features, discussed in the above, are in good agreement
with the experimental observation for a non-interacting spin-orbit
coupled atomic Fermi gas; see, for example, Fig. 4 in Ref. \cite{Wang2012S}
and Fig. 2 in Ref. \cite{Cheuk2012S}. 

\subsubsection{The rf spectroscopy of the molecular component}

\begin{figure}
\includegraphics[width=1\columnwidth]{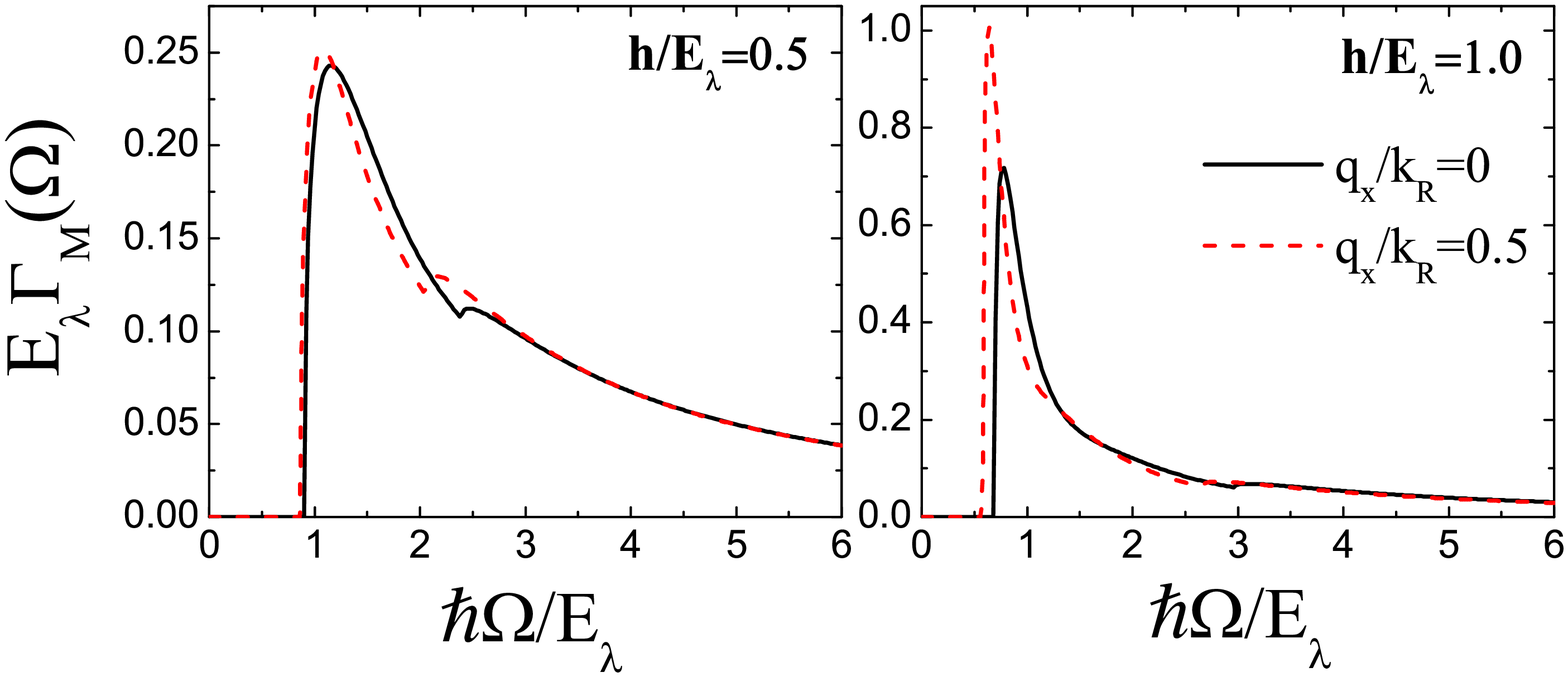}

\caption{(Color online) The rf spectroscopy of a single molecule for different
values of the Zeeman field at given COM momenta $\mathbf{q}/k_{R}=0$
and $\mathbf{q}/k_{R}=0.5$. Here, we have chosen an interaction strength
$E_{B}/E_{\lambda}=1$.}

\label{fig5}
\end{figure}

With the two-body binding energy solved from Eq. (\ref{eq:3.2.16})
at a given COM momentum $\mathbf{q}$, the rf transition strength
of a single molecule can be calculated directly from Eq. (\ref{eq:4.3.6}).
We plot the rf response of a single molecule with zero COM momentum,
as well as non-zero COM momentum (i.e., $q_{x}/k_{R}=0.5$) in Fig.
\ref{fig5}. Here, the interatomic interaction strength, denoted by
$E_{B}/E_{\lambda}=\left[\hbar^{2}/\left(m\lambda a_{s}\right)\right]^{2}$,
is set to be unity. Since the SO coupling is only along the $x$ direction,
we can set the transverse COM momentum $\mathbf{q}_{\perp}=0$, which
is physically irrelevant and arbitrary. From Fig. \ref{fig5}, we
find that there is a slight red-shift in the peak position with a
non-zero COM momentum. This is because the energy gap between the
final and initial states becomes smaller at non-zero COM momentum,
and consequently a little higher transition peak appears. 

\begin{figure}
\includegraphics[width=1\columnwidth]{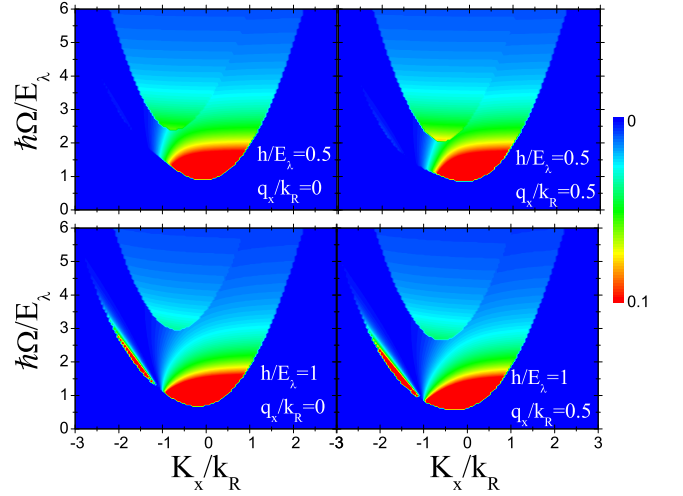}

\caption{(Color online) The momentum-resolved rf spectroscopy of a single molecule
for different values of the Zeeman field at given COM momenta $\mathbf{q}_{x}/k_{R}=0$
and $\mathbf{q}/k_{R}=0.5$. Here, we have chosen the interaction
strength $E_{B}/E_{\lambda}=1$.}

\label{fig6}
\end{figure}

The momentum-resolved rf spectroscopy of a single molecule can be
obtained from Eq. (\ref{eq:4.3.6}) by integrating over $\mathbf{k}_{\perp}$,
and is given by
\begin{multline}
\Gamma_{M}\left(\Omega,k_{x}\right)=\frac{m}{8\pi^{2}\hbar\mathcal{C}}\left[\left|s_{\mathbf{q}/2+\mathbf{k}+}\right|^{2}\Theta\left(k_{\perp,+}^{2}\right)+\right.\\
\left.\left|s_{\mathbf{q}/2+\mathbf{k}-}\right|^{2}\Theta\left(k_{\perp,-}^{2}\right)\right],\label{eq:4.4.18}
\end{multline}
 where $\Theta\left(x\right)$ is the step function and
\begin{multline}
k_{\perp,\pm}^{2}=\frac{m}{\hbar}\left[\Omega+\varepsilon_{B}\left(\mathbf{q}\right)\right]-\left[\frac{k_{x}^{2}+\left(k_{x}+k_{R}\right)^{2}}{2}+\right.\\
\left.\frac{k_{R}^{2}+\left(k_{R}+q_{x}\right)^{2}}{4}\pm\frac{m}{\hbar^{2}}\sqrt{h^{2}+\lambda^{2}\left(\frac{q_{x}}{2}+k_{x}\right)^{2}}\right].\label{eq:4.4.19}
\end{multline}
 The momentum-resolved rf spectroscopy of a single molecule with zero
and non-zero COM momentum ($q_{x}/k_{R}=0.5$) are presented in Fig.
\ref{fig6}. In this plot, we have defined $K_{x}=-k_{x}-k_{R}$,
as we require that in the absence of SO coupling the spectra is an
even function of $K_{x}$ \cite{Hu2012R}. We find that the contribution
from two final states are well separated in different frequency domain
and therefore should be easily observed experimentally.

\subsubsection{The total rf spectroscopy of the gas mixture of atoms and molecules}

By increasing the attractive interaction strength of a two-component
Fermi gas at finite temperatures, the system will evolve into a mixture
of fermionic atoms and bosonic molecules. We have so far discussed
the integrated and momentum-resolved rf responses for both atoms and
molecules. The total rf spectroscopy of a harmonically trapped\emph{
}ideal\emph{ }gas mixture of atoms and molecules can now be easily
obtained.

For the atomic component, the rf transition strength is already given
by Eq. (\ref{eq:4.4.14}). As to the molecules, the population $N_{M}$
can be calculated from Eq. (\ref{eq:2.2.11}) at given temperature
$T/T_{F}$ and interaction strength $E_{B}/E_{\lambda}$. In real
experiments, the temperature is low so that the distribution of the
molecular COM momentum $\mathbf{q}$ is very narrow around $\mathbf{q}=0$.
As already seen from Figs. 5 and 6, the rf spectroscopy of molecules
depends weakly on $\mathbf{q}$. Thus, as a good qualitative approximation,
we may focus on the rf transition of molecules with zero COM momentum
only. Then, the total rf transition strength of molecular component
is given by $N_{M}\Gamma_{M}\left(\Omega\right)$. In the end, we
obtain the total rf spectroscopy,
\begin{equation}
\Gamma_{t}\left(\Omega\right)=\Gamma_{A}\left(\Omega\right)+N_{M}\Gamma_{M}\left(\Omega\right).\label{eq:4.4.20}
\end{equation}

\begin{figure}
\includegraphics[width=1\columnwidth]{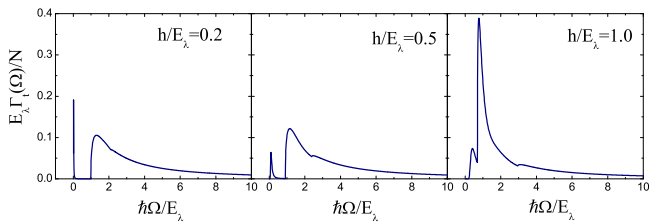}

\caption{(Color online) The total rf spectroscopy of a harmonically trapped
ideal gas mixture of fermionic atoms and bosonic molecules at different
Zeeman field. Here, we take $E_{B}/E_{\lambda}=1$, $E_{F}/E_{\lambda}=1$
and $T/T_{F}=0.2$ .}

\label{fig7}
\end{figure}

Fig. \ref{fig7} shows the total rf spectroscopy of the atom-molecule
mixture at different Zeeman field and at fixed temperature $T/T_{F}=0.2$
and interaction strength $E_{B}/E_{\lambda}=1$. It is obvious that
there is a double-peak structure in the spectroscopy. The peak at
higher frequency is responsible for weakly bound molecules, since
more energy is required for pair breaking. With increasing the Zeeman
field $h$, the energy difference between the final and initial states
of atoms during the rf transition becomes larger, thus a decrease
in the atomic peak is observed. On the other hand, since the energy
difference between the molecular final and initial states becomes
smaller as the Zeeman field increases, the rf signal of molecules
grows, as we may anticipate. 

\begin{figure}
\includegraphics[width=1\columnwidth]{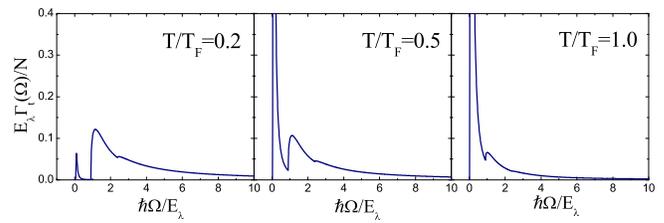}

\caption{(Color online) The total rf spectroscopy of a harmonically trapped
ideal gas mixture of fermionic atoms and bosonic molecules as a function
of temperature at a fixed Zeeman field strength $h/E_{\lambda}=0.5$.
Here, we use $E_{B}/E_{\lambda}=1$ and $E_{F}/E_{\lambda}=1$. }

\label{fig8}
\end{figure}

In order to show the temperature dependence, we also calculate the
total rf response of the ideal gas mixture at different temperatures
and at a fixed Zeeman field $h/E_{\lambda}=0.5$, as reported in Fig.
\ref{fig8}. As the temperature decreases, the population of molecules
becomes larger. Thus, a reduction of the atomic signal is observed.
The molecular response should dominate when temperature decreases
to zero.

\begin{figure}[b]
\includegraphics[width=1\columnwidth]{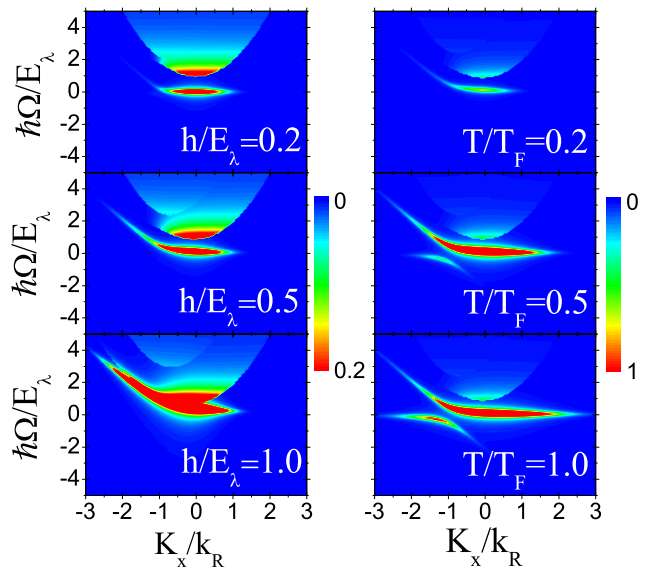}

\caption{(Color online) The total momentum-resolved rf spectroscopy of a harmonically
trapped ideal gas mixture of fermionic atoms and bosonic molecules.
The left panel shows the spectra at different Zeeman field strength
at a fixed temperature $T/T_{F}=0.2$, while the right panel shows
the temperature dependence of spectroscopy at a fixed Zeeman field
$h/E_{\lambda}=0.5$. Here, we take $E_{B}/E_{\lambda}=1$ and $E_{F}/E_{\lambda}=1$.
Note the different color scale in the left and right panels.}

\label{fig9}
\end{figure}
Finally, we present the total momentum-resolved rf spectroscopy in
Fig. \ref{fig9}, with parameters correpsonding to Fig. \ref{fig7}
(see the left panel) and Fig. \ref{fig8} (right panel).

\section{conclusions\label{sec:conclusions}}

In conclusion, we have theoretically investigated the momentum-resolved
radio-frequency spectroscopy of a harmonically trapped\emph{ }ideal\emph{
}gas mixture of fermionic atoms and bosonic molecules with spin-orbit
coupling. This is a \emph{qualitative} theoretical model of spin-orbit
coupled atomic Fermi gases near broad Feshbach resonances. 

In particular, for local uniform systems, general formulas for the
wavefunction and binding energy of weakly bound bosonic molecules
with \emph{non-zero} center-of-mass momentum have been derived. The
influence of this non-zero center-of-mass momentum on the radio-frequency
spectroscopy of molecules has been discussed in detail. Based on few-body
solutions in uniform systems, the radio-frequency responses of atomic
and molecular components in harmonic traps have been calculated within
local density approximation, respectively, by using Fermi's golden
rule. Our results provide a qualitative picture of momentum-resolved
radio-frequency spectroscopy of a strongly interacting spin-orbit
coupled Fermi gas in harmonic traps, which could be directly observed
in future experiments at Shanxi University \cite{Wang2012S,ZhangPrivateComm}
and MIT\cite{Cheuk2012S,ZwierleinPrivateComm}.

In real experiments, where strongly interacting Fermi gases are created
in the vicinity of Feshbach resonances, the atoms and weakly bound
molecules must strongly interact with each other. An in-depth analysis
of radio-frequency spectroscopy would rely on much more complicated
many-body theories \cite{Hu2006}, beyond the simple few-body picture
presented in this work.
\begin{acknowledgments}
Shi-Guo Peng and Kaijun Jiang are supported by China Postdoctoral
Science Foundation (Grant No. 2012M510187), the NSFC projects (Grant
No. 11004224 and No.11204355) and the NFRP-China (Grant No. 2011CB921601).
Xia-Ji Liu and Hui Hu are supported by the ARC Discovery Projects
(Grant No. DP0984637 and No. DP0984522) and the NFRP-China (Grant
No. 2011CB921502). \end{acknowledgments}

\end{document}